\documentclass[onecolumn,preprintnumbers]{revtex4}
\usepackage{makeidx}
\usepackage{amssymb}
\usepackage{amsmath}
\usepackage{mathrsfs}
\usepackage{graphicx}
\usepackage{dcolumn}
\usepackage{bm}
\usepackage[center]{subfigure}
\usepackage{color}

\begin{document}

\title{Hybrid matter-wave - microwave solitons on the lattice}
\author{Zhihuan Luo$^1$, Weiwen Luo$^1$, Wei Pang$^2$, Zhijie Mai$^1$,
Yongyao Li$^{3}$}
\email{yongyaoli@gmail.com}
\author{Boris A. Malomed$^{4,1}$}
\email{malomed@post.tau.ac.il}
\affiliation{$^{1}$ Department of Applied Physics, South China Agricultural University,
Guangzhou 510642, China \\
$^{2}$ Department of Experiment Teaching, Guangdong University of
Technology, Guangzhou 510006, China \\
$^{3}$ School of Physics and Optoelectronic Engineering, Foshan University,
Foshan $528000$, China \\
$^{4}$ Department of Physical Electronics, School of Electrical Engineering,
Faculty of Engineering, and the Center for Light-Matter Interaction, Tel
Aviv University, Tel Aviv 69978, Israel}

\begin{abstract}
We introduce a two-component system which models a pseudospinor
Bose-Einstein condensate (BEC), with a microwave field coupling its two
components. The feedback of BEC of the field (the local-field effect) is
taken into account by dint of the respective Poisson equation, which is
solved using the Green's function. This gives rise to an effective
long-range self-trapping interaction, which may act alone, or be combined
with the contact cubic nonlinearity. The system is made discrete by loading
the BEC into a deep optical-lattice potential. Numerical solutions
demonstrate that onsite-centered fundamental solitons are stable in the
cases of attractive or zero contact interactions, while offsite-centered
solitons are unstable. In the case of the repulsive onsite nonlinearity,
offsite solitons are stable, while their onsite-centered counterparts are
stable only at sufficiently small values of the norm, where bistability
between the off- and onsite-centered mode takes place. The shape of the
onsite-centered solitons is very accurately predicted by a variational
approximation (which includes essential technical novelties).
Spatially-antisymmetric (\textquotedblleft twisted") solitons are stable at
small values of the norm, being unstable at larger norms. In the strongly
asymmetric version of the two-component system, which includes the Zeeman
splitting, the system is reduced to a single discrete Gross-Pitaevskii
equation, by eliminating the small higher-energy component.
\end{abstract}

\maketitle

\section{Introduction}

It is well known that a pair of couterpropagating coherent optical waves
produce an effective dipole potential for cold atoms, in the form of an
optical lattice (OL). OL-induced potentials play a well-known role in
dynamics of quantum atomic gases, helping to create gap solitons and other
species of collective modes \cite{VVK,Morsch}. The feedback of the atomic
motion on the light propagations, alias the local-field effect (LFE),
produces a deformation of the OL. The LFE is often weak, due to the far
detuning of the OL's frequency from the atomic resonance. Nevertheless, in
some situations, such as the diffraction of Bose-Einstein condensates (BECs)
on optical waves with unequal intensities \cite{Li-2008}, LFE is a
significant factor. In particular, it can make the underlying OL relatively
\textquotedblleft soft", inducing an effective nonlocal interaction \cite%
{Zhu-2011}, which, in turn, can create matter-wave solitons, even in the
absence of contact interactions in the BEC \cite{Dong-2013}.

Another manifestation of the LFE is a nonlinear correction to the coupling
of two components of the wave function of a pseudospinor BEC, corresponding
to two different hyperfine atomic states, via the magnetic component of the
microwave (MW) field which resonantly couples these states \cite%
{Minhang1,Minhang2} (this is the difference from earlier studied cases, in
which the electric component played a dominant role in the action of the
LFE). The LFE-induced nonlinear correction is nonlocal in this case, as the
magnetic field obeys the respective Poisson equation. The final result is a
system of two Gross-Pitaevskii equations (GPEs) coupled by an effective
nonlinear interaction, which is locally repulsive but globally attractive,
thus opening a new way to create self-trapped hybrid matter-wave - microwave
solitons.

It is well known that BECs, split into a chain of droplets by a deep OL
potential, are adequately modeled, in the framework of the tightly-binding
approximation for the droplets, by discrete GPEs. In this approximation,
hopping between adjacent sites of the underlying lattice is represented by
linear couplings between them \cite{tight1}-\cite{tight5}. If atoms in BEC\
carry permanent magnetic moments, the respective discrete GPE features
long-range interactions between remote sites \cite{Sandra,Pfau}, which, in
particular, gives rise to stable on- and off-site centered solitons and
periodically modulated extended patterns. In a more general context,
discrete GPEs belong to the broad class of discrete nonlinear Schr\"{o}%
dinger equations (NLSEs), which appear in a great variety of physical
realizations \cite{review,Kevr}. In this context, nonlocal interactions are
also a topic of considerable interest \cite{Gaididei}.

The objective of the present work is to introduce a discrete version of the
coupled GPE system with the strongly nonlocal nonlinearity, which was
introduced in the continuum form in work \cite{Minhang1}. It implies that
the two-component BEC, resonantly coupled by the microwave field, is loaded
in a deep OL potential, that leads to the effective discretization of the
system. In addition to the physical significance of the model, it provides a
novel variety of discrete NLSEs, with a specific long-range inter-site
interaction. The system gives rise to specific on- and offsite-centered
fundamental spatially even discrete solitons, as well as \textit{twisted}
\cite{Jena,twisted,review} (spatially antisymmetric) ones. These soliton
families are also quite different from those in the original continuum
system, that were reported in work \cite{Minhang1}. In particular, the width
of the discrete fundamental solitons may be much smaller in comparison with
their counterparts in the continuum limit.

The rest of the paper is organized as follows. The model is introduced in
section II, which is followed by the presentation of numerical and
analytical findings in sections II and III, respectively. In particular,
analytical results for the fundamental on-site-centered solitons are
obtained by means of a variational approximation (VA), which very accurately
match the numerical findings, this version of the VA for discrete solitons
being an essential technical novelty by itself. Further, the system with
asymmetry between its components, induced by detuning (Zeeman splitting)
between them, is studied in section IV. The paper is concluded by section V.

\section{The model}

We consider the one-dimensional hybrid system introduced in \cite{Minhang1}.
It includes the pseudospinor (two-component) BEC, with wave functions of two
hyperfine states coupled by the magnetic component of the microwave field.
The feedback of the pseudospinor wave function on the microwave field,
represented by the Poisson equation for the field, coupled to the system of
two GPEs, which include the field, is a manifestation of the LFE in the
present setting. The system is made discrete by fragmentation in a deep OL\
potential. As a result, the microwave field can be eliminated, solving the
one-dimensional Poisson equation by means of the Green's function (similarly
to how it was done in \cite{Minhang1}). As a result, the system is modeled
by coupled discrete GPEs for amplitudes $({\Psi }_{n},{\Phi }_{n})$ of the
pseudospinor wave function at OL sites, with discrete coordinate $n$:
\begin{eqnarray}
&&i\frac{\partial {\Psi }_{n}}{\partial t}=-\frac{1}{2}\left( {\Psi }_{n+1}-2%
{\Psi }_{n}+{\Psi }_{n-1}\right) +\eta {\Psi }_{n}-H_{0}{\Phi }_{n}+\gamma {%
\Phi }_{n}\sum_{m}|m-n|{\Phi }_{m}^{\ast }{\Psi }_{m}+\left( G\left\vert
\Psi _{n}\right\vert ^{2}+\Gamma \left\vert \Phi _{n}\right\vert ^{2}\right)
\Psi _{n},  \notag \\
&&i\frac{\partial {\Phi }_{n}}{\partial t}=-\frac{1}{2}\left( {\Phi }_{n+1}-2%
{\Phi }_{n}+{\Phi }_{n-1}\right) -\eta {\Phi }_{n}-H_{0}{\Psi }_{n}+\gamma {%
\Psi }_{n}\sum_{m}|m-n|{\Psi }_{m}^{\ast }{\Phi }_{m}+\left( G\left\vert {%
\Phi }_{n}\right\vert ^{2}+\Gamma \left\vert \Psi _{n}\right\vert
^{2}\right) {\Phi }_{n}.  \label{eqn-model}
\end{eqnarray}%
In the general case, the system includes the self- and cross- contact
interactions with coefficients $G$ and $\Gamma $, respectively (alias the
onsite nonlinearity, in terms of the lattice dynamics). The onsite
interactions are repulsive, in the case of $G,\Gamma >0$, and attractive,
for $G,\Gamma <0$. Further, the coefficient of the intersite hopping is
scaled to be $1$ in Eq. (\ref{eqn-model}), $\eta $ (which may be zero) is
the strength of the Zeeman-splitting effect \cite{Zeeman}, which induces the
detuning of two-component system, $H_{0}$ is the background magnetic field
of the microwave, which induces linear mixing between the two components.
The terms $\sim \gamma $ represent the long-range interaction generated by
the magnetic field, the particular form of the interaction kernel in Eq. (%
\ref{eqn-model}) being produced by the above-mentioned Green's function used
for the solution of the one-dimensional Poisson equation.

Stationary states are looked for in the usual form,
\begin{equation}
({\Psi }_{n}(t),{\Phi }_{n}(t))=(\psi _{n},\phi _{n})e^{-i\mu t},
\label{Psipsi}
\end{equation}%
where $(\psi _{n},\phi _{n})$ are stationary wave functions, and $\mu $ is a
real chemical potential. The norms of the two components are defined as
\begin{equation}
N_{\psi }=\sum_{n}|\psi _{n}|^{2},N_{\phi }=\sum_{n}|\phi _{n}|^{2}.
\label{eqn-norm}
\end{equation}%
The relative norm difference between $\psi _{n}$ and $\phi _{n}$, which is
induced by the Zeeman splitting, when it is present, is
\begin{equation}
\Delta =\frac{|N_{\psi }-N_{\phi }|}{N}.  \label{eqn-norm-diff}
\end{equation}%
Widths of the two components of the discrete soliton are defined as
\begin{equation}
W_{\psi }=\frac{\left( \sum_{n}|\psi _{n}|^{2}\right) ^{2}}{\sum_{n}|\psi
_{n}|^{4}},W_{\phi }=\frac{\left( \sum_{n}|\phi _{n}|^{2}\right) ^{2}}{%
\sum_{n}|\phi _{n}|^{4}}.  \label{width}
\end{equation}

First, it is relevant to produce the dispersion relation for the linearized
version of Eq. (\ref{eqn-model}), substituting the solution in the form of $%
\left( {\Psi }_{n}(t),{\Phi }_{n}(t)\right) \sim \exp \left( -i\mu
t+iqn\right) $ with real wavenumber $q$:%
\begin{equation}
\mu =2\sin ^{2}\left( q/2\right) \pm H_{0}.  \label{mu}
\end{equation}%
However, unlike models with the local nonlinearity, where bright solitons
may be found solely in bandgaps of the spectrum, i.e., in intervals of the
chemical potential\ which cannot be covered by values given by Eq. (\ref{mu}%
) with arbitrary real values of $q$, this commonly known principle does not
apply to the present system, because the nonlocal nonlinear term in Eq. (\ref%
{eqn-model}) grows $\sim |n|$ at $|n|\rightarrow \infty $, making the system
\emph{nonlinearizable}, as concerns the identification of the asymptotic
shape of the soliton's shapes, i.e., necessary conditions admitting the
existence of localized modes \cite{Minhang1} (see also works \cite%
{Barcelona1,Barcelona2}, as concerns more general classical and quantum
discrete systems which admit bright solitons in nonlinearizable models with
the repulsive nonlinearity).

\section{The symmetric system (no Zeeman splitting)}

\subsection{The formulation}

The long-range interaction in the hybrid system plays a crucial role in the
creation of solitons. To reveal its nature, we first consider the basic case
of zero-detuning, with $\eta =0$ in Eq. (\ref{eqn-model}). In this case, the
equation for symmetric states, with $\Phi _{n}=\Psi _{n}$, reduces to
\begin{equation}
i\frac{\partial \Psi _{n}}{\partial t}=-\frac{1}{2}\left( \Psi _{n+1}-2\Psi
_{n}+\Psi _{n-1}\right) -H_{0}\Psi _{n}+\gamma \Psi _{n}\sum_{m}|m-n||\Psi
_{m}|^{2}+\left( G+\Gamma \right) \left\vert \Psi _{n}\right\vert ^{2}\Psi
_{n}.  \label{eqn-symm}
\end{equation}
By means of an obvious transformation,
\begin{equation}
{\Psi }_{n}(t)=\gamma ^{-1/2}e^{iH_{0}t}{U}_{n}(t),  \label{transform}
\end{equation}%
one may cast Eq. (\ref{eqn-symm}) in the form of
\begin{equation}
i\frac{\partial U_{n}}{\partial t}=-\frac{1}{2}\left(
U_{n+1}-2U_{n}+U_{n-1}\right) +U_{n}\sum_{m}|m-n||U_{m}|^{2}+g\left\vert
U_{n}\right\vert ^{2}U_{n},  \label{eqn-symm-simple}
\end{equation}%
which contains the single free parameter, $g\equiv \left( G+\Gamma \right)
/\gamma $.

Then, stationary solutions with real chemical potential are looked for as
\begin{equation}
U_{n}(t)=e^{-i\mu t}u_{n},  \label{Uu}
\end{equation}%
where discrete field $u_{n}$ satisfies the equation
\begin{equation}
\mu u_{n}+\frac{1}{2}\left( u_{n+1}-2u_{n}+u_{n-1}\right)
=u_{n}\sum_{m}|m-n|u_{m}^{2}+gu_{n}^{3}.  \label{u}
\end{equation}

\subsection{The variational approximation (VA)}

The usual approach to constructing discrete solitons in nonlinear lattice
models starts from the anti-continuum limit, which neglects the hopping
between adjacent sites \cite{Aubry,Kevr} (in the continuum counterpart of
the lattice model, it corresponds to the Thomas-Fermi approximation \cite{TF}%
). However, the application of the anti-continuum limit to Eq. (\ref{u})
does not make the remaining equation solvable, because it includes the
nonlocal interaction.

A more promising analytical approach may be based on the VA. To this end, we
note that Eq. (\ref{u}) can be derived from the Lagrangian,%
\begin{equation}
L=-\frac{\mu }{2}N+H,  \label{L}
\end{equation}%
where the respective norm and Hamiltonian are%
\begin{gather}
N=\sum_{n=-\infty }^{+\infty }u_{n}^{2},  \label{norm} \\
H=\frac{1}{4}\left\{ \sum_{n=-\infty }^{+\infty }\left[ \left(
u_{n+1}-u_{n}\right) ^{2}+gu_{n}^{4}\right] +\sum_{m,n=-\infty }^{+\infty
}\left\vert m-n\right\vert u_{m}^{2}u_{n}^{2}\right\} ~.  \label{Ham}
\end{gather}%
Following the pattern of works \cite{Weinstein}-\cite{Van}, the variational
ansatz for fundamental onsite-centered discrete solitons, with amplitude $A$
and inverse width $a$, can be adopted as
\begin{equation}
u_{n}=A\exp \left( -a|n|\right) .  \label{ans}
\end{equation}%
Norm (\ref{norm})\ of the ansatz is%
\begin{equation}
N=A^{2}\coth a.  \label{N}
\end{equation}

The application of the VA to Lagrangian (\ref{L})-(\ref{Ham}), which
includes the long-range interaction, is a novel technical problem. Its
solution is facilitated by the use of the following formulas, which appear
in the course of the substitution of ansatz (\ref{ans})\ in Lagrangian (\ref%
{L}):
\begin{equation}
\sum_{n=-\infty }^{+\infty }e^{-2a|n|}=\coth a,\sum_{n=-\infty }^{+\infty
}\left( e^{-a\left\vert n+1\right\vert }-e^{-a\left\vert n\right\vert
}\right) ^{2}=2\tanh \left( \frac{a}{2}\right) ,  \label{sum1}
\end{equation}%
\begin{equation}
\sum_{m,n=-\infty }^{+\infty }\left\vert m-n\right\vert e^{-2a\left(
|m|+|n|\right) }\equiv 2\sum_{m\geq n=-\infty }^{+\infty }\left( m-n\right)
e^{-2a\left( |m|+|n|\right) }=\frac{2\cosh \left( 2a\right) +1}{2\sinh
\left( 2a\right) \cdot \sinh ^{2}a}.  \label{sum2}
\end{equation}%
Then, the effective Lagrangian corresponding to ansatz (\ref{ans}) is
explicitly calculated as%
\begin{gather}
L_{\mathrm{eff}}=-\frac{\mu }{2}A^{2}\coth a+\frac{A^{2}}{2}\tanh \left(
\frac{a}{2}\right) +\frac{gA^{4}}{4}\coth \left( 2a\right)   \notag \\
+\frac{A^{4}}{8}\frac{2\cosh \left( 2a\right) +1}{\sinh \left( 2a\right)
\cdot \sinh ^{2}a}~.  \label{eff}
\end{gather}%
Finally, the variational equations are derived from the Lagrangian as%
\begin{equation}
\frac{\partial L_{\mathrm{eff}}}{\partial a}=\frac{\partial L_{\mathrm{eff}}%
}{\partial \left( A^{2}\right) }=0.  \label{VA}
\end{equation}%
After straightforward manipulations, Eq. (\ref{VA}) can be cast in the form
of expressions for the amplitude and chemical potential in terms of the
inverse width:
\begin{gather}
A^{2}=\frac{4\sinh (2a)\cdot \sinh ^{3}a}{(1-g)[\cosh (4a)+2]+3g\cosh (2a)},
\label{A} \\
\mu =1-\cosh a+\frac{\sinh ^{2}a}{\cosh a}\cdot \frac{2g\sinh ^{2}a+8\cosh
^{4}a+1}{(1-g)[\cosh (4a)+2]+3g\cosh (2a)}.  \label{mu-VA}
\end{gather}%
These prediction are compared to numerical findings in the next subsection.

\subsection{Numerical results}

\subsubsection{Single-soliton states}

\begin{figure}[tbp]
\includegraphics[scale=0.32]{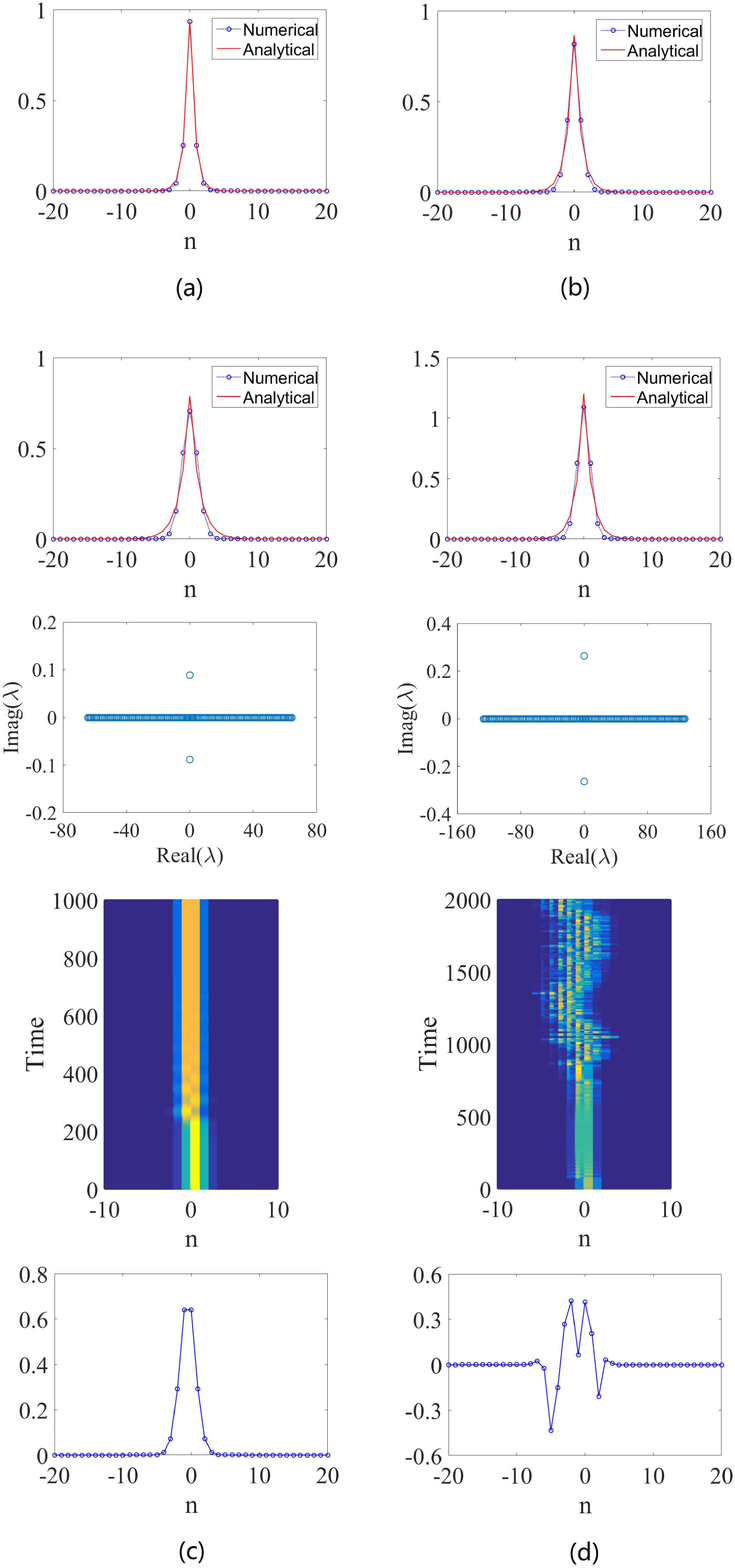}
\caption{(Color online) Typical examples of fundamental onsite-centered
solitons in the symmetric system (with zero detuning, $\protect\eta =0$),
along with the analysis of their stability, as produced by Eqs. (\protect\ref%
{eqn-symm-simple}) and (\protect\ref{u}). The analytical results, shown by
red solid lines, are produced by the variational approximation, based on
Eqs. (\protect\ref{ans}) and (\protect\ref{A}), (\protect\ref{mu-VA}), while
the numerical findings are represented by blue circles. Parameters of stable
solitons are $g=-1$, $N=1$ in (a), and $g=0$, $N=1$ in (b). For unstable
solitons, the parameters are $g=1$, $N=1$ in (c), and $g=1$, $N=2$ in (d).
In (c), the unstable onsite-centered soliton spontaneously transforms into a
stable offsite-centered one, which is shown in the bottom panel. In (d), the
unstable soliton develops a turbulent pattern. The difference in the
outcomes of the instability development correlates with the fact that the
instability growth rate, Im$\left( \protect\lambda \right) $, is larger in
(d).}
\label{fig1-fundamental}
\end{figure}
\begin{figure}[tbp]
\includegraphics[scale=0.28]{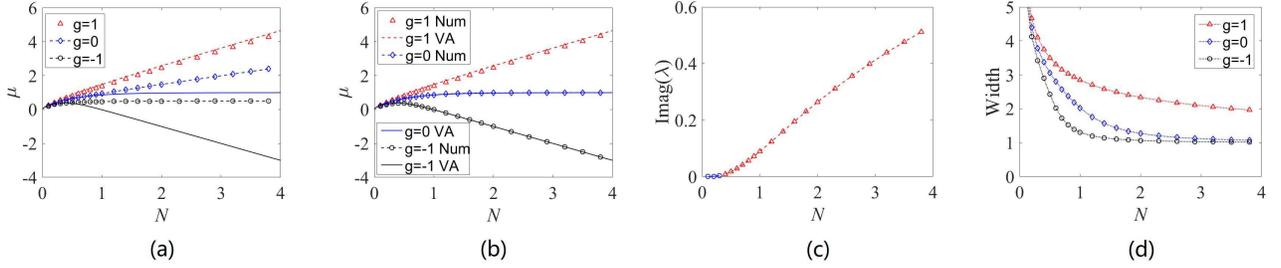}
\caption{(Color online) (a) and (b): Dependences $\protect\mu (N)$ for
offsite- and onsite-centered discrete solitons at different values of $g$ in
the symmetric system ($\protect\eta =0$). Red triangles, blue diamonds, and
black circles present numerical results produced by Eq. (\protect\ref%
{eqn-symm-simple}) for $g=1$, $0$, and $-1$, respectively. The corresponding
variational results are shown in (b) by red dashed, blue, and black solid
lines, respectively. (c) Onsite-centered solitons: The instability growth
rate of the onsite-centered branch with $g=1$ versus the total norm, $N$.
The solitons are stable and unstable in regions covered by blue circles and red triangles,
respectively, while the branches with $g=-1$ and $0$ are
completely stable. For the offsite-centered solitons, the situation is
opposite: they are completely stable for $g=1$, and completely unstable for $%
g=-1$ and $0$. (d) The width of the onsite-centered solitons, defined as per
Eq. (\protect\ref{width}), versus $N$.}
\label{fig2-norm}
\end{figure}

\begin{figure}[tbp]
\includegraphics[scale=0.32]{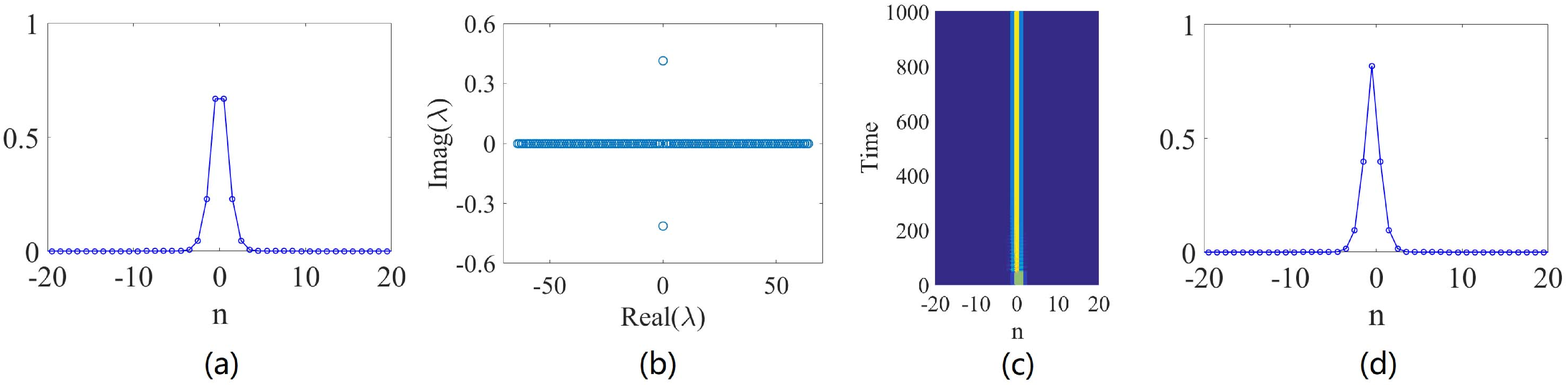}
\caption{(Color online) A typical examples of unstable offsite-centered
solitons in the symmetric system with $g=0$ and $N=1$. Eventually, it
transforms into a stable onsite-centered one, which is shown in panel (d). }
\label{fig2-off-onsite}
\end{figure}

\begin{figure}[tbp]
{\includegraphics[width=.98\columnwidth]{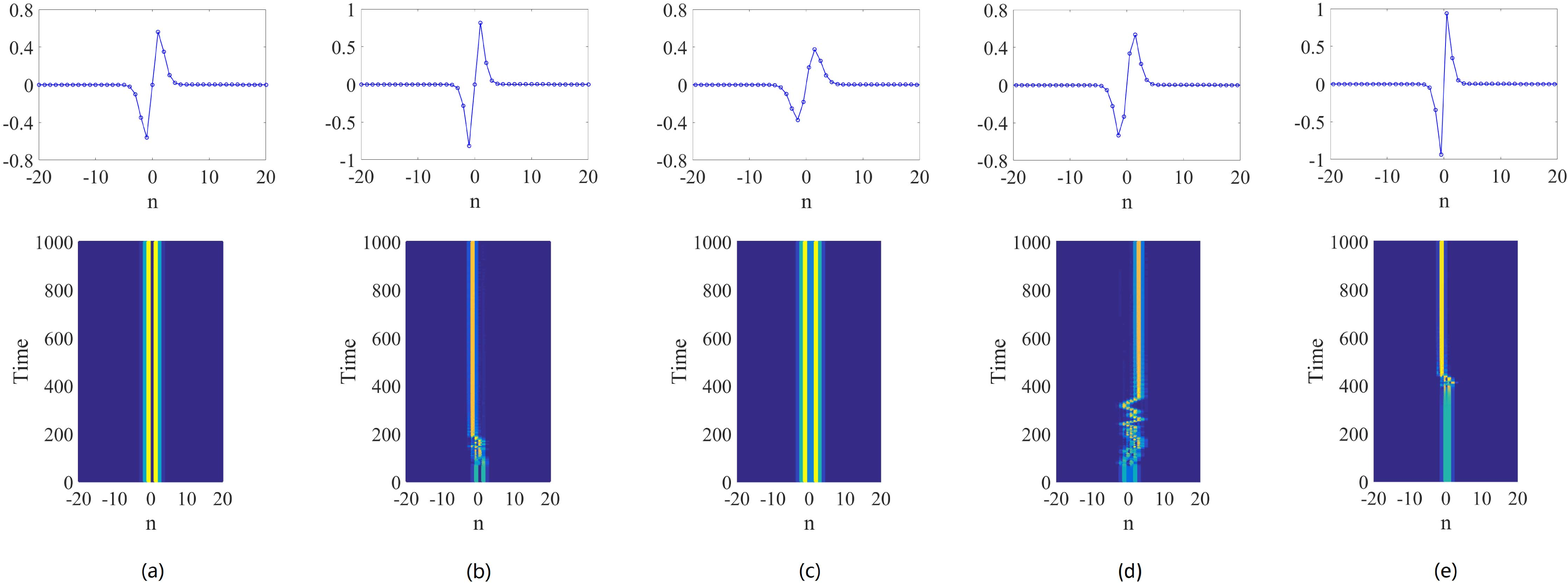}}
\caption{(Color online) Typical examples of twisted solitons and their
evolution in the symmetric system ($\protect\eta =0$). (a) and (b):
Onsite-centered solitons with $N=0.9$ and $N=1.5$. (c), (d) and (e):
Offsite-centered solitons with $N=0.5$, $N=0.9$ and $N=2$, respectively. The
twisted solitons are stable in (a) and (c), and unstable in other cases. In
all plots, $g=0$ (no onsite (contact) nonlinearity). }
\label{fig3-excited}
\end{figure}

Typical examples of fundamental onsite-centered solitons in the
zero-detuning system ($\eta =0$), produced by numerical solution of Eq. (\ref%
{eqn-symm-simple}), along with their analytical counterparts provided by the
VA, are demonstrated in Fig. \ref{fig1-fundamental}. Close agreement between
the numerical and variational results is evident. Stability of the discrete
solitons was checked both in direct simulations of their perturbed
simulations, and through computation of eigenfrequencies, $\lambda $, for
small perturbations governed by the linearization of Eq. (\ref%
{eqn-symm-simple}), the instability growth rate, if any, being Im$\left(
\lambda \right) $. The solitons are completely stable in the cases of $g=-1$
and $g=0$, i.e., in the cases of the attractive or zero onsite nonlinearity
in Eq. (\ref{eqn-symm-simple}), independently of the soliton's norm $N$. In
case of $g=+1$, i.e., repulsive onsite interactions, solitons become
unstable at sufficiently large $N$, as shown in Figs. \ref{fig1-fundamental}%
(c,d). The bottom panels in (c) and (d) demonstrate the shape of the
discrete field observed after long evolution.

We summarize results for the family of such solitons in the form of the
dependence of their chemical potential, $\mu $, and width, $W$, on the norm,
$N$. In the case of $g=0$, as seen in Fig. \ref{fig2-norm}(b), $\mu (N)$ is
a monotonously increasing function with a positive slope, $d\mu /dN>0$, thus
satisfying the \textit{anti-Vakhitov-Kolokolov (VK)\ criterion}, which is a
necessary stability condition for bright solitons supported by repulsive
nonlinearities (e.g., gap solitons) \cite{anti} (the VK criterion per se, $%
d\mu /dN<0$, is necessary for the stability of solitons created by an
attractive nonlinearity \cite{VK1,VK2}). On the other hand, in the case of
the competition of the nonlocal and contact interactions ($g\neq 0$) the
anti-VK and VK criteria may be invalid.

We conclude that the onsite solitons belonging to the branches with $g=0\ $%
and $-1$ are stable for all values of the total norm. On the contrary, in
the case of $g=1$ (the self-repulsive onsite nonlinearity), the onsite
solitons are stable only at $N\leq 0.4$, being unstable at $N>0.4$. The
analytical predictions produced by the VA, denoted by dashed lines in Fig. %
\ref{fig2-norm}(b), are in excellent agreement with their numerical
counterparts.

The branches of offsite-centered fundamental solitons, which are displayed
in Fig. \ref{fig2-norm}(a), feature a totally different behavior. The
analysis of their stability readily shows that the branch corresponding to $%
g=1$ is \emph{completely} \emph{stable}, while ones with $g=-1$ and $0$ are
\emph{completely unstable}. Thus, the system demonstrates \emph{bistability}
in the case of $g=1$ and $N\leq 0.4$, where both offsite- and
onsite-centered discrete solitons are stable. In the latter case, the
calculation of the respective values of Hamiltonian (\ref{Ham}) demonstrates
that the solitons of the onsite type realize a smaller value of $H$, i.e..,
they represent the ground state.

Lastly, Fig. \ref{fig2-norm}(d) demonstrates that the discrete solitons
shrink under the action of stronger nonlinearity, which corresponds to
larger $N$. Additional simulations demonstrate that unstable
offsite-centered solitons tend to spontaneously transform into their stable
counterparts of the onsite type, as shown in Fig. \ref{fig2-off-onsite}.

The first excited states in the form of twisted discrete solitons have also
been found in the symmetric system, as demonstrated in Fig. \ref%
{fig3-excited}. These states are stable when norm $N$ is small, e.g., at $%
N=0.8$ for the onsite-centered state, and at $N=0.5$ for the offsite one,
see panels (a) and (c), respectively. At larger $N$, after a sufficiently
long evolution the twisted mode eventually evolves into a stable fundamental
soliton, which is true for both the onsite and offsite modes, as seen in
panels (b), (d) and (e) in Fig. \ref{fig3-excited}. We have identified a
critical value of the norm, $N_{c}$, above which the twisted states lose
their stability: $N_{c}\simeq 0.96$ and $N_{c}\simeq 0.59$ for the on- and
offsite-centered ones, associated with the twisted patterns containing one
or two intermediate sites, respectively (see Fig. \ref{fig3-excited}). Thus,
the twisted state of the onsite-centered type is more stable than its
offsite counterpart. The number of intermediated sites between two maxima of
opposite signs in the twisted state depends on the norm. The larger the norm
is, the fewer the number of intermediated sites becomes. In the case of the
onsite-centered type, we can find a tightest-shaped twisted mode with only
one intermediate site, as shown in Fig. \ref{fig3-excited}(a). Twisted modes
of the offsite type contain, as least, two intermediated sites, see Fig. \ref%
{fig3-excited}(c). For sufficient large $N$, one can find the tightest
offsite-centered twisted state which has no intermediate sites. However, it
is found to be unstable, as shown in Fig. \ref{fig3-excited}(e).

\begin{figure}[tbp]
{\includegraphics[width=0.8\columnwidth]{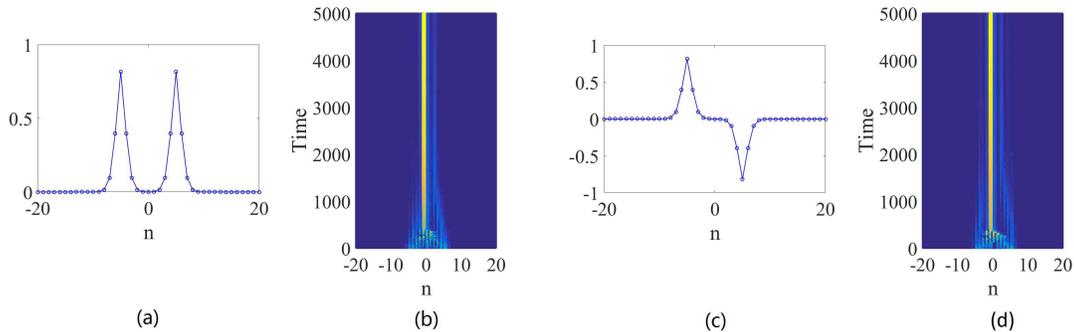}}
\caption{(Color online) Typical examples of bound states of two fundamental
discrete solitons in the symmetric system ($\protect\eta =0$). (a) and (b):
The in-phase bound state and its evolution. (c) and (d): The out-of-phase
bound state and its evolution. In all plots, the distance between two peaks
is $10$. }
\label{fig4-phase}
\end{figure}

\subsubsection{Bound states of fundamental solitons}

We have also constructed bound states of two fundamental solitons of
in-phase and out-of-phase types, i.e., with identical and opposite signs of
the two constituents, respectively, and with different separations between
them. Such bound states, associated with initial separations, are always
unstable, finally evolving into single fundamental discrete solitons, as
shown in Fig. \ref{fig4-phase}. Naturally, the time for the spontaneous
transformation of the in-phase (or out-of-phase) states into fundamental
solitons is longer if the initial distance between two peaks is larger.

\begin{figure}[tbp]
\includegraphics[scale=0.3]{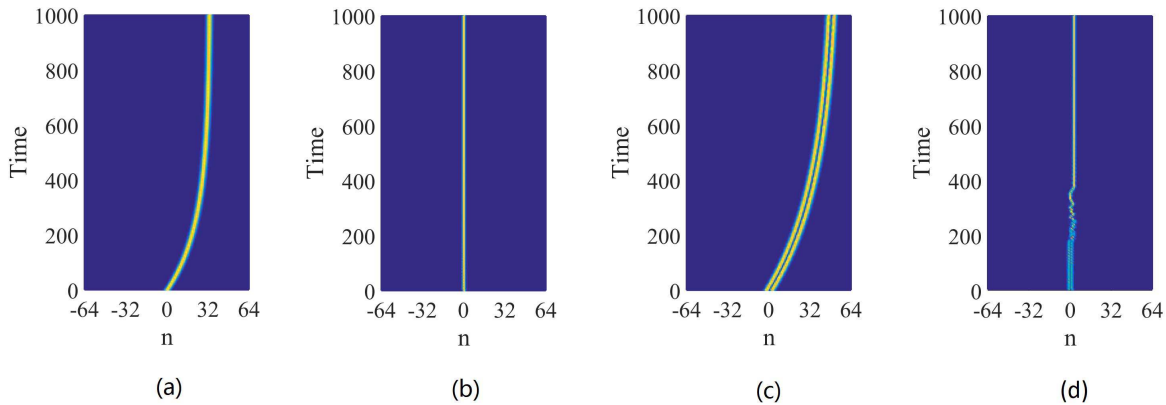}
\caption{(Color online) Discrete solitons kicked as per Eq. (\protect\ref%
{t=0}), with $k=0.05\protect\pi $. (a) and (b): Fundamental solitons with $%
N=0.2$ and $N=1$. (c) and (d): Dipoles with $N=0.2$ and $N=1$. }
\label{fig5-kick}
\end{figure}

\subsubsection{Mobility of the discrete solitons}

It is well known that, with the exception of integrable systems, such as the
Toda \cite{TL} and Ablowitz-Ladik \cite{AL} lattices, and some other
specially designed models \cite{Barash}, rigorous solutions for moving
discrete solitons do nor exist. Nevertheless, simulations of the discrete
NLSE with the onsite cubic nonlinearity, initiated by a kick, $k$, applied
to a quiescent discrete soliton, $u_{n}$, i.e.,
\begin{equation}
U_{n}(t=0)=\exp \left( ikn\right) u_{n},  \label{t=0}
\end{equation}%
demonstrate its robust mobility: it keeps moving through the lattice
indefinitely long, without visible emission of radiation waves
(\textquotedblleft phonons") or loss of the velocity \cite{mobility1}-\cite%
{mobility5}.

To address the possible mobility of the solitons in the present system, we
have performed simulations of Eq. (\ref{eqn-model}) with initial conditions (%
\ref{t=0}). At small $N$, the fundamental soliton sets in motion, but
eventually it comes to a halt, as shown in Fig. \ref{fig5-kick}(a). When $N$
is sufficiently large, the kicked soliton does not feature even transient
mobility, remaining quiescent, as shown in Fig. \ref{fig5-kick}(b). Thus,
the present system is drastically different in terms of the soliton's
mobility from the discrete NLSE with the local cubic term.

The mobility of twisted discrete solitons was studied too. It was found that
their dynamics is similar to that of the fundamental ones when $N$ is small
(e.g., for $N=0.2$), i.e., the kicked twisted state features transient
mobility. However, when $N$ increases to $1$, the kick destroys the twisted
state, in contrast to the situation for the robust fundamental solitons.

\begin{figure}[tbp]
\includegraphics[scale=0.32]{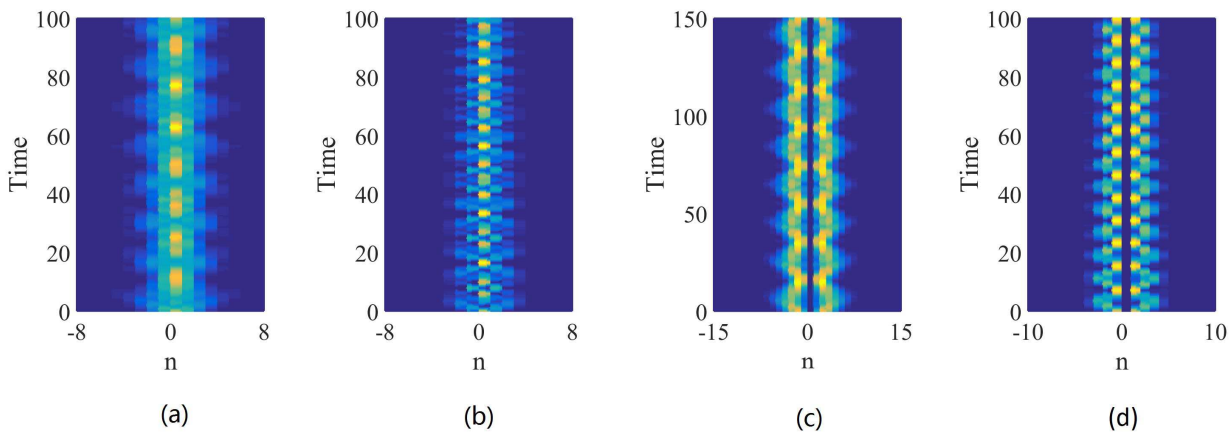}
\caption{(Color online) The evolution of discrete solitons to which chirp $b$
is applied. (a) and (b): Fundamental solitons with $N=0.2$, $b=0.1\protect%
\pi $ and $N=1$, $b=0.5\protect\pi $. (c) and (d): Twisted solitons with $%
N=0.2$, $b=0.1\protect\pi $ and $N=1$, $b=0.5\protect\pi $. }
\label{fig6-chirp}
\end{figure}

In addition, it is possible to simulate effects of real \textit{chirp} $b$
applied to the fundamental and twisted discrete solitons, by solving Eq. (%
\ref{eqn-model}) with initial condition
\begin{equation}
U_{n}(t=0)=\exp \left( ib|n|\right) u_{n}.  \label{b}
\end{equation}%
The results displayed in Fig. \ref{fig6-chirp} demonstrate excitation of
robust internal modes in both the fundamental and twisted solitons as the
reaction to the application of the chirp.

\section{The asymmetric (detuned) system}


\subsection{An analytical approximation for strong asymmetry}

To address the general form of Eq. (\ref{eqn-model}), which includes
detuning $\eta $, it is relevant to start from the limit case of large $\eta
$ (strong detuning), when the approximation which was developed, for the
continuum systems, in works \cite{Dong-2013} and \cite{Sherman} suggests
that this case may be effectively reduced to a single equation, by
eliminating the small higher-energy component in favor of the large
lower-energy one. To this end, the underlying wave functions are redefined as%
\begin{equation}
\left( \Psi _{n},\Phi _{n}\right) \equiv e^{i\eta t}\left( \tilde{\Psi}%
_{n}(t),\tilde{\Phi}_{n}(t)\right) .  \label{tilde}
\end{equation}%
Substituting this in Eq. (\ref{eqn-model}) for $\Psi _{n}$ leads, in the
first approximation, to relation%
\begin{equation}
\tilde{\Psi}_{n}=\left( H_{0}/2\eta \right) \tilde{\Phi}_{n}.  \label{PsiPhi}
\end{equation}%
Next, the substitution of this relation in the equation for $\Phi _{n}$
produces an equation which, up to a shift of the chemical potential and
different notation for the coefficient, is tantamount to Eq. (\ref{eqn-symm}%
):%
\begin{equation}
i\frac{\partial \tilde{\Phi}_{nn}}{\partial t}=-\frac{1}{2}\left( \tilde{\Phi%
}_{n+1}-2\tilde{\Phi}_{n}+\tilde{\Phi}_{n-1}\right) -\frac{H_{0}^{2}}{2\eta }%
\tilde{\Phi}_{n}+\frac{\gamma H_{0}^{2}}{4\eta ^{2}}\tilde{\Phi}%
_{n}\sum_{m}|m-n||\tilde{\Phi}_{m}|^{2}+\Gamma \left\vert \tilde{\Phi}%
_{n}\right\vert _{n}^{2}\tilde{\Phi}.  \label{Phi-tilde}
\end{equation}%
Thus, the limit case of strong asymmetry between the two components leads to
essentially the same single GPE as in the case of the fully symmetric system.

\subsection{Numerical results}

In the presence of finite detuning, a typical example of a stable
two-component discrete soliton with components $\psi _{n}$ and $\phi _{n}$
(see Eq. (\ref{Psipsi}) along with simulations of its evolution, initiated
by the addition of small random perturbations, is displayed in Fig. \ref%
{fig7-wave}.

\begin{figure}[tbp]
\centering
\includegraphics[scale=0.35]{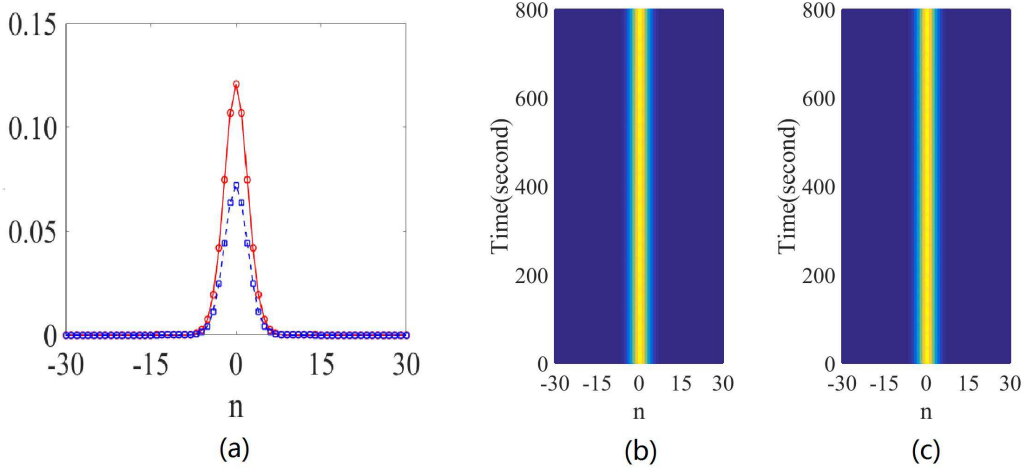}
\caption{(Color online) (a) A typical example of the two components $\protect%
\phi $ and $\protect\psi $\ of the fundamental discrete soliton (red circles
and blue squares, respectively) in the asymmetric (mismatched) system. (b)
and (c): The perturbed evolution of the $|\Psi |$ and $|\Phi |$ components,
corroborating stability of this soliton. The parameters are $N=1$, $\protect%
\eta =0.5$, $H_{0}=2$, and $\protect\gamma =0.1$.}
\label{fig7-wave}
\end{figure}

\begin{figure}[tbp]
\centering
\includegraphics[scale=0.5]{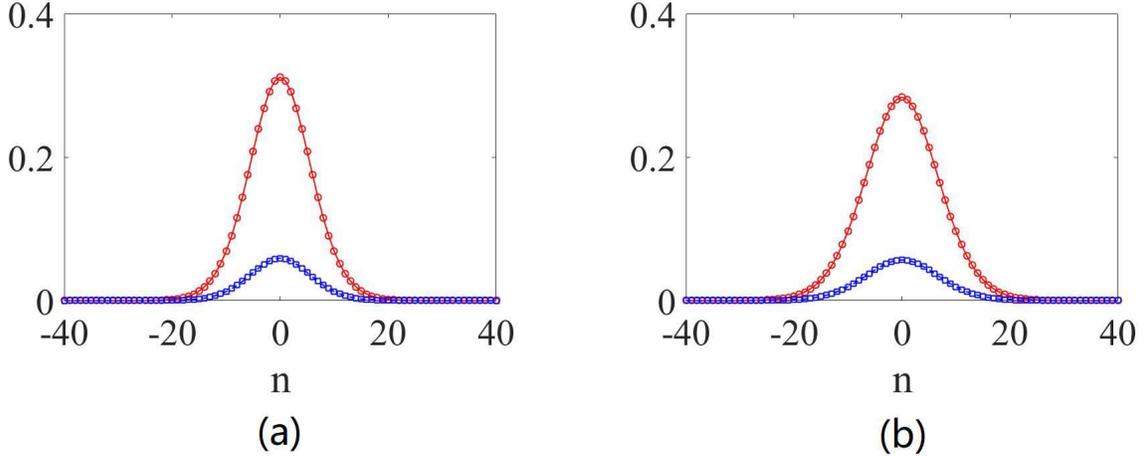}
\caption{(Color online) (a) Typical examples of components $\protect\phi _{n}
$ and $\protect\psi _{n}$ (blue circles and black squares, respectively)\ of
the fundamental discrete soliton in the strongly asymmetric system, as
produced by Eq. (\protect\ref{eqn-model}). Red diamonds and blue stars
depict the corresponding approximation produced by Eqs. (\protect\ref{PsiPhi}%
) and (\protect\ref{Phi-tilde}). The parameters are $\protect\eta =10$, $%
H_{0}=4$, $\protect\gamma =0.1$, and $N=1$. }
\label{fig8-strongdetuning}
\end{figure}

In the case of large detuning, the full numerical solution for a stable
fundamental discrete soliton is compared to the approximation, based on Eqs.
(\ref{PsiPhi}) and (\ref{Phi-tilde}), in Fig. \ref{fig8-strongdetuning}. It
is seen that the approximation is accurate for the strongly asymmetric
solitons.

\begin{figure}[tbp]
\includegraphics[scale=0.35]{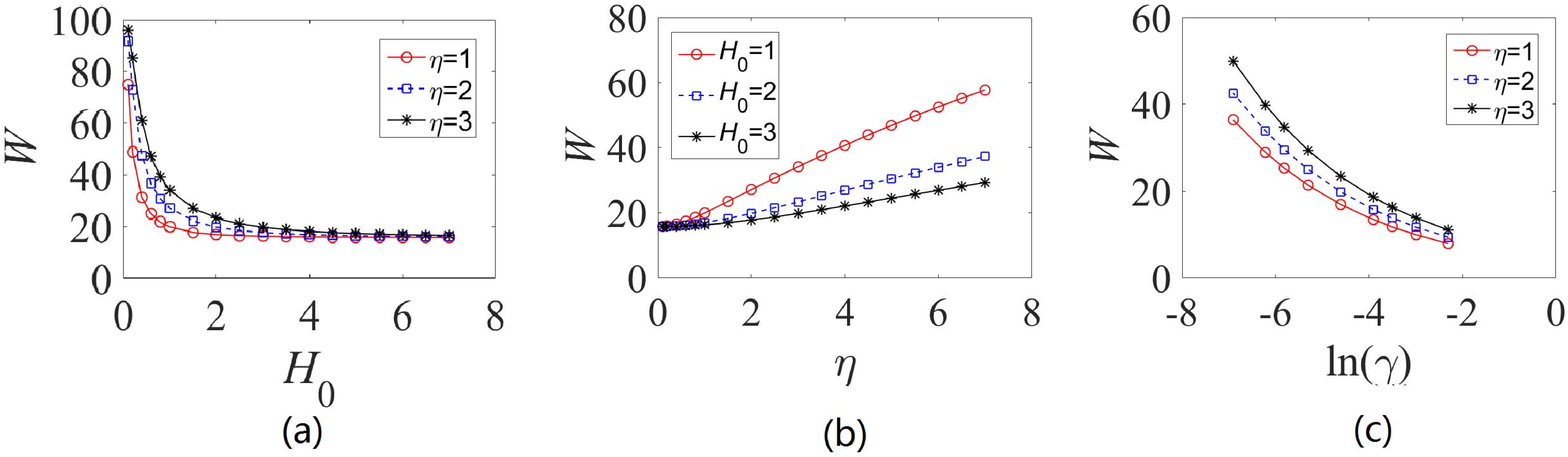}
\caption{(Color online) (a) The width of fundamental discrete solitons
versus background magnetic field $H_{0}$ for a fixed value of the strength
of the long-range interaction, $\protect\gamma =0.01$, and different values
of the Zeeman-splitting coefficient, $\protect\eta $. (b) The width versus $%
\protect\eta $ for $\protect\gamma =0.01$ and different values of $H_{0}$.
(c) The width versus $\ln (\protect\gamma )$ for $H_{0}=2.0$ and different
values of $\protect\eta $. In all plots, $N=1$.}
\label{fig9-width-detuning}
\end{figure}

The width of the fundamental discrete solitons is plotted, as a function of
the background magnetic field, the strength of the Zeeman splitting, and the
strength of the long-range interaction, in Fig. \ref{fig9-width-detuning}.
It is seen that the increase of the background field and strength of the
long-range interaction leads to shrinkage of the solitons. On the contrary,
stronger Zeeman splitting makes the solitons wider. Note that the width of
the fundamental discrete solitons diverges in the limit of $H_{0}\rightarrow
0$, while, in the case of $\eta =0$, the solution does not depend on $H_{0}$%
, as seen in panel (b).

Lastly, to consider the distribution of the total norm (which is now scaled
to be $N=1$) between the two components ($\psi ,\phi $) of the asymmetric
discrete solitons, a typical example of the dependence of the component
norms on the background magnetic field is shown in Fig. \ref%
{fig10-norm-detuning}(a). One can see that the difference between the lower-
and higher-energy components decreases as $H_{0}$ increases, because the
effect of the Zeeman splitting is suppressed by the background field. The
dependence of the norms on $\eta $ is shown in panel (b), which naturally
implies that the Zeeman splitting enhances the asymmetry.

\begin{figure}[tbp]
\includegraphics[scale=0.35]{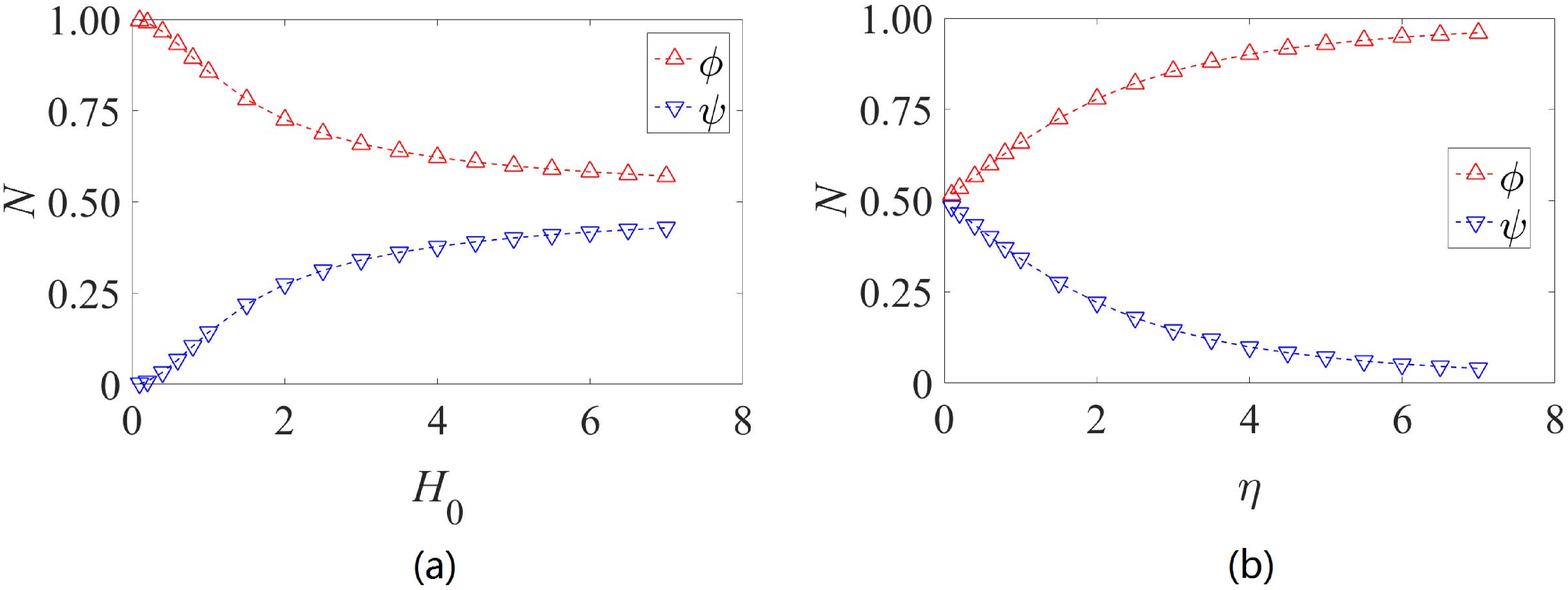}
\caption{(Color online) (a) A typical example of norms of the $\protect\psi $
and $\protect\phi $ components as functions of $H_{0}$ for $\protect\eta =1$%
. (b) The dependence of the norms on $\protect\eta $ for $H_{0}=3$. In all
plots, $N=1$ and $\protect\gamma =0.01$. }
\label{fig10-norm-detuning}
\end{figure}

\section{Conclusion}

We have introduced a two-component discrete system which models the
pseudospinor BEC, whose components are coupled by the microwave field, while
the condensate is made effectively discrete by a deep optical-lattice
potential. The elimination of the field by means of the Green's function of
the one-dimensional Poisson equation gives rise to a long-range
self-trapping interaction between lattice sites, which can be combined with
the usual onsite cubic nonlinearity of either sign. The numerical solution
shows that onsite-centered fundamental solitons are stable in the case of
zero and attractive onsite interactions, while offsite-centered fundamental
solitons are unstable. The situation is opposite in the case of the
repulsive onsite nonlinearity: solitons of the offsite type are stable,
while onsite-centered ones are stable only at sufficiently small values of
the norm, at which the system features the bistability of the onsite and
offsite-centered fundamental solitons. The variational approximation very
accurately predicts the shape of the onsite-centered solitons. The first
excited states, in the form of twisted solitons, are stable at small values
of the norm, while at large norms they are subject to an instability which
transforms them into fundamental solitons. Bound states of two fundamental
solitons of both in-phase and out-of-phase types are unstable, evolving into
a single fundamental-like soliton. Another specific feature of the system is
that it does not admit mobility of the discrete solitons.

In the asymmetric version of the system, the background magnetic field tends
to compress the solitons and suppress the asymmetry between the lower- and
higher- components, while the Zeeman splitting plays an opposite role. In
the limit case of strong asymmetry, the two-component system can be
approximately reduced to a single discrete GPE (Gross-Pitaevskii equation),
which is corroborated by numerical results.

A challenging direction for further work is to introduce a two-dimensional
version of the system and construct two-dimensional discrete solitons in it,
including discrete vortices.

\section{Acknowledgements}

YL acknowledges the supports of the National Natural Science Foundation of
China (Grants Nos. 11874112 and 11575063). The work of BAM on this topic is
supported, in part, by grant No. 1287/17 from the Israel Science Foundation.
This author appreciates hospitality of the Department of Applied Physics at
the South China Agricultural University.

\end{document}